\documentclass[12pt,preprint]{aastex}

\shorttitle{Magnetic Landscape of Sun's Polar Region}
\shortauthors{Tsuneta et al.}

\begin{document}

\title{Magnetic Landscape of Sun's Polar Region}

\author{S. Tsuneta\altaffilmark{1}, K. Ichimoto\altaffilmark{1}, Y. Katsukawa\altaffilmark{1},
B. W. Lites\altaffilmark{2}, K. Matsuzaki\altaffilmark{3}, S. Nagata\altaffilmark{4}, 
D. Orozco Suarez\altaffilmark{5}, T. Shimizu\altaffilmark{3}, M. Shimojo\altaffilmark{6},
R. A. Shine\altaffilmark{7}, Y. Suematsu\altaffilmark{1}, T. K. Suzuki\altaffilmark{8},
T. D. Tarbell\altaffilmark{7}, A. M. Title\altaffilmark{7}}

\altaffiltext{1}{National Astronomical Observatory, Mitaka, Tokyo, 181-8588, Japan}
\altaffiltext{2}{High Altitude Observatory, National Center for Atmospheric 
Research\altaffilmark{9}, P.O. Box 3000, Boulder CO 80307-3000, USA}
\altaffiltext{3}{ISAS/JAXA, Sagamihara, Kanagawa, 229-8510, Japan}
\altaffiltext{4}{Kwasan and Hida Observatories, Kyoto University, Yamashina, Kyoto, 607-8471, Japan}
\altaffiltext{5}{Instituto de Astrofisica de Andalucia (CSIC), Camino Bajo de Huetor, 50  18008 Granada - Spain}
\altaffiltext{6}{Nobeyama Solar Radio Observatory, National Astronomical Observatory of Japan, Nobeyama, Nagano, 284-1305, Japan}
\altaffiltext{7}{Lockheed Martin Solar and Astrophysics Laboratory, B/252, 3251 Hanover St., Palo Alto, CA 94304, USA}
\altaffiltext{8}{School of Arts and Sciences, University of Tokyo, 3-8-1, Komaba, Meguro, Tokyo 153-8902}
\altaffiltext{9}{The National Center for Atmospheric Research
        is sponsored by the National Science Foundation.}

\begin{abstract}
We present the magnetic landscape of the polar region of the Sun that is unprecedented in terms of high spatial resolution, large field of view, and polarimetric precision. These observations were carried out with the Solar Optical Telescope aboard \emph{Hinode}. Using a Milne-Eddington inversion, we found many vertically-oriented magnetic flux tubes with field strength as strong as 1 kG that are scattered in latitude between $70\,^{\circ}$ $\sim$ $90\,^{\circ}$. They all have the same polarity, consistent with the global polarity of the polar region. The field vectors were observed to diverge from the center of the flux elements, consistent with a view of magnetic fields that expand and fan out with height. The polar region is also covered with ubiquitous horizontal fields. The polar regions are the source of the fast solar wind channelled along unipolar coronal magnetic fields whose photospheric source is evidently rooted in the strong field, vertical patches of flux.  We conjecture that vertical flux tubes with large expansion around the photosphere-corona boundary serve as efficient chimneys for Alfv\'en waves that accelerate the solar wind. 
\end{abstract}


\keywords{sun: magnetic fields, polar region, solar wind, dynamo}

\section{Introduction}
The Sun's polar magnetic fields are thought to be the direct manifestation of the global poloidal fields in the interior, which serve as seed fields for the global dynamo that produces the toroidal fields responsible in active regions and sunspots. The polar regions are also the source of fast solar winds. Although the polar regions are of crucial importance to the dynamo process and acceleration of the fast solar winds, its magnetic properties are poorly known. Magnetic field measurements in the solar polar regions have long been a challenge: variable seeing combined with the strong intensity gradient and the foreshortening effect at the solar limb greatly increases the systematic noise in ground-based magnetographs. Nevertheless, pioneering observations have been carried out for the polar regions \citep{tw90, lvz94, o04a, of04, hkm97, br07}. These observations typically have provided only the measurements of the line-of-sight magnetic component. Full Stokes polarimetry has also been carried, but as with most of the ground-based observations described above, the spatial resolution of those measurements was limited by seeing \citep{bw96}. Another limitation of the past polar observations is that they have been restricted to individual polar faculae within a small field of view, and have not provided us with a global magnetic landscape of the polar region except for \emph{GONG/SOLIS} \citep{jh07}. 

We investigated the properties of photospheric magnetic field in polar regions using the Solar Optical Telescope, SOT \citep{st07, ys07, ki07, ts07}. SOT is a diffraction-limited (0".2\,-\,0".3) Gregorian telescope with filter-graph and spectro-polarimeter aboard the satellite \emph{Hinode} \citep{tk07}. These observations are unprecedented in terms of their very high spatial resolution, wide field of view, and high polarimetric sensitivity and accuracy in measurements of vector magnetic fields.

\section{Stokes maps of the polar region}
\emph{Hinode} observed the polar regions on March 16, 2007, when the South pole was located $7\,^{\circ}$ inside the visible solar disk. Stokes profiles of the two FeI lines (630.2 and 630.3 nm) were observed with the spectro-polarimeter (SP) of SOT. Degrees of linear (Stokes Q and U) and circular (Stokes V) polarization are defined by $\frac{\int_{}^{}\mathstrut Q (\lambda)d\lambda}{I_{c}\int_{}^{}\mathstrut d\lambda}$, $\frac{\int_{}^{}\mathstrut U(\lambda)d\lambda}{I_{c}\int_{}^{}\mathstrut d\lambda}$, $\frac{\int_{}^{}|V(\lambda)|d\lambda}{I_{c}\int_{}^{}\mathstrut d\lambda}$, respectively, where $I_{c}$ is the continuum intensity, and $Q, U, V$ are the observed Stokes profiles. The SP wavelength sampling is 2.15pm. When the red lobe of the Stokes V profile is positive, a minus sign is added, meaning that the line-of-sight field is directed away from the observer. The integration is performed between  $-$21.6pm and $-$4.32pm and between $+$4.32pm and $+$21.6pm from the center of individual Stokes I profiles. These wavelength ranges are determined by examining the actual data. If the range is too wide, the maps will be more susceptible to the photon noise. Maps of these degrees of polarization are hereafter called the Stokes Q, U and V maps, and are shown in Figure 1. These data were taken at 12:02:19-14:55:48 on March 16, 2007. 

Position angles for the Stokes Q and U were determined as follows: 0 degree is to the East-West, 90 degree to the North-South. If the transverse field is directed to the West (North), it will be seen white (black) in the Stoke Q map. If the transverse field is directed 45 degree (135 degree), it will be seen white (black) in the Stoke U map. Black (plus polarity) in the Stokes V map indicates magnetic field is directed toward the observer, and white (minus polarity) away from the observer.

We noticed scattered isolated patches in the Stokes Q map (Figure 1(b)). They coherently have minus sign (black), indicating the presence of a magnetic component vertical (North-South) to the local surface. These patches are associated with a bipolar structure in the Stokes U map (Figure 1(c)). Figure 1(d) and (e) show the Stokes V map. The same Stokes V image seen from just above the south pole is shown in Figure 2, where the foreshortening effect has been corrected. All the pixels in the observing coordinate (as shown in Figure 1) are mapped to the corresponding positions in the polar maps (Figures 2 and 3), and the gaps in between the mapped pixels are interpolated. Figures 1(d) and 2 show a bipolar structure corresponding to the patches seen in the Stokes Q and U maps in higher latitudes. An example is indicated by a circle in Figures 1 and 2.
The bipolar structure in the Stokes V map together with the Q and U maps suggests fanning-out flux tubes vertical to local surface. 

Unipolar patches with plus (black) polarity are dominant at lower latitudes in the Stokes V map. This is because the vertical flux tubes located at lower latitude have all the fanning-out magnetic vectors directed toward the observer. The dominant black polarity indicates a component of the field oriented toward the observer. This shows that all of the vertical magnetic patches are coherently directed away from the Sun, consistent with the general polar fields in 2007 \citep{ng03, du04}.

We also noticed in the Stokes U map that the plus (white) polarity in the bipolar structures is more dominant in the East side, while minus (black) polarity in the West side. Since the flux tubes vertical to the local surface in the East side is tilted toward the plus axis of the Stokes U, the bipolar structures in the East side are more plus-biased (white), and vice versa. This explains East-West asymmetry of the bipolar structures in the Stokes U map.

\section{Polar landscape} 
\subsection{Milne-Eddington fitting}

We applied a least-squares fitting to the Fe I 630.15~nm and Fe I
630.12~nm observed Stokes profiles assuming a Milne-Eddington
atmosphere with the MILOS code \citep{os07, dos07}. The 10
free parameters are: the three components describing the vector
magnetic field (strength $B$, inclination angle $\gamma$, and azimuth
angle $\chi$), the line of sight velocity, two parameters describing
the source function, the line-to-continuum absorption coefficient
ratio, the Doppler width, the damping parameter, and the stray light
factor $\alpha$. To minimize the influence of noise, we have analyzed
only pixels whose polarization signal peaks exceed a given threshold
above the noise level $\sigma$. The noise level was determined in the
continuum wavelength range of the profiles. The fitting is performed
for pixels whose $Q$, $U$ or $V$ signals are larger than $5 \sigma$.
It turns out that 10.5\% of the area meets the criteria.
 
There may be unresolved magnetic elements along the line of sight, e.g.,
rays passing obliquely through vertical flux tubes observed close to
the limb. A non-zero stray-light factor $\alpha$ may be interpreted 
as a parameter including both the filling factor of a non- or 
weakly-magnetized atmosphere along the line of sight and the stray-light
contamination factor. The stray-light profile is evaluated
individually for each pixel as the average of the Stokes $I$ profiles
observed in a 1\arcsec-wide box centered on the pixel \citep{dos07}. 
This arrangement also allows us to accurately estimate the stray-light
profiles of rapidly changing continuum intensity toward the limb.
If the stray-light contamination is negligible then the effective 
filling factor of the magnetic atmosphere will be represented
by $f=1-\alpha$. 

The results of our inversions applied to the polar region
show that the distribution of the effective filling factors has a
broad peak at $f = 0.15$ with FWHM range $0.05 < f < 0.35$ 
(Figure 5 (d)). Orozco Suarez et al.(2007b) suggested that there is 
considerable stray-light contribution, and actual filling factor may 
be larger than the nominal values derived above. In an extreme case, 
we estimate the magnetic flux by assuming $f=1$ in the subsequent
sections. 

\subsection{Vertical kilo-Gauss patch and horizontal field}
Figure 3 is a map of the magnetic field strength as seen from just above the south pole. Such polar representation is needed to correctly see the spatial extent and size distribution of the magnetic islands in the polar region. While many of them are isolated, and some have the form of a chain of islands, complex internal structures are seen inside the individual patches. Many patchy magnetic islands have very high field strength reaching above 1 kG. These kG patches coincide in position with those seen in Figure 1: They are coherently unipolar, and like plage and network fields at lower latitudes \citep{mp97}, they have magnetic field vertical to the local surface. The fanning-out structure is confirmed in the vector magnetic field map resulting from the least-squares fitting.

We noticed a clear tendency for patches to be larger in size with increasing latitude. The size is as large as 5"$\times$5" at higher latitudes and 1"$\times$1" at lower latitudes. Degradation in spatial resolution due to the projection effect may contribute to the larger size at high latitude. This, however, implies that close to the solar limb, we observe flux tubes higher in the atmosphere. The response function to temperature \citep{dti04} for the core of the Stokes I profile has a broad peak between 100km and 500km above continuum optical depth unity in Sun center observations. The response function for a plane-parallel atmosphere viewed obliquely at an angle of $80\,^{\circ}$ has a peak that is $50\,-\,100\,$km higher, implying that we observe higher altitude atmosphere at $80\,^{\circ}$. 

We are interested in the inclination angle $i$ of the magnetic field vector with respect to the local normal. Close to the limb, it is possible to determine the inclination $i$ of the magnetic field vector with respect to the local surface without the usual $180\,^{\circ}$-ambiguity \citep{dti04} of the transverse field components: the inclination angle is derived by $\cos i = \cos \gamma \cos \theta + \sin \gamma \sin \theta \cos (\pi/2-\chi)$, where $\theta$ is the latitude, $\gamma$ the inclination of magnetic field vector, and $\chi$ the azimuth angle. 
Figure 4 shows the inclination of the field lines: Red contours indicate regions where the local inclination $i$ smaller than than $25\,^{\circ}$ (vertical), while blue contours show regions with local inclination larger than $65\,^{\circ}$ (horizontal). All the large patches have fields that are vertical (red) to the local surface, while the smaller patches tend to be horizontal (blue). Most of the magnetic structures seen in Figure 3 thus have either vertical or horizontal directions. These two types do not appear to be spatially correlated. 

Figure 4 also indicates that magnetic patches of larger spatial size coincide in position with polar faculae \citep{lvz94, o04a, of04}. Indeed, Figure 5 (c) shows a histogram of the continuum intensity relative to the local average intensity for pixels of magnetic field strength larger than 300G and 800G. The distribution of the horizontal fields is essentially symmetric about average intensity, while the vertical fields tend to have higher continuum intensities. These bright points correspond to the polar faculae. Vertical kG patches change considerably in shape and distribution in $5\,-\,10\,$hours, while the spatial distribution of horizontal magnetic field changes completely in $30\,$min or less. This is consistent with the observations of the horizontal fields made by  \emph{SOLIS} \citep{jh07} and  \emph{Hinode} \citep{rc07, ri07}. 
 
We have selected 41 patches with vertical magnetic field manually to derive the distribution of magnetic flux and patch size as a function of latitude. The total magnetic flux of a magnetic patch is estimated by $\sum_{i} B_{i} f_{i} s_{i}$, where $B_{i}, f_{i}$ are the intrinsic magnetic field strength and the filling factor of the ${i}$-th SOT pixel inside the patch respectively, and $s_{i}$ the common pixel size. The magnetic flux of the patches ranges from $1.8 \times 10^{18}$Mx to $1.0 \times 10^{20}$Mx with mean flux of $2.7 \times 10^{19}$Mx. We obtain the size of the magnetic patches $\sum_{i} s_{i}$, where the summation is done for pixels with per-pixel average field strength $B_{i} f_{i}$ larger than 10 G: The size increases by a factor of 1.92, and the flux $\sum_{i} B_{i} f_{i} s_{i}$ decreases by a factor of 1.33 between latitudes $70\,^{\circ}$ and $85\,^{\circ}$. The total magnetic flux is preserved within 40\%.

Figure 5(a) shows the histogram (probability distribution function) of the magnetic field strength $B$ for latitudes $>75\,^{\circ}$: vertical magnetic fields with inclination $i < 25\,^{\circ}$  dominate the stronger field regime, while horizontal fields with $i > 65\,^{\circ}$ are much more prevalent below 250 G. 41\% of the pixels for which the inversion are performed were occupied by vertical magnetic field, and 49\% horizontal magnetic field. Figure 5 (a) can be compared with PDF [figure 7 of Orozco Suarez et al.(2007b)] for the quiet sun obtained with \emph{Hinode}. A magnetic energy PDF defined by the number of pixels multiplied by $B^2$ as a function of field strength is shown in Figure 5 (b). This shows where the magnetic energy is mainly located as a function of field strength. The vertical flux tubes with higher field strength are energetically dominant, while weaker horizontal flux tubes contrastingly carry more energy. 

\subsection{Total magnetic flux in polar region} 

The total \emph{vertical} magnetic flux in the SOT field of view is $2.2 \times 10^{21}$ Mx, while the total \emph{horizontal} flux is $4.0 \times 10^{21}$ Mx. The effective filling factor is taken into account to estimate these total fluxes: The flux of an individual pixel in Figure 3 is estimated by $B \times f \times$pixel size with foreshortening correction. 

Considering the stray-light contribution, the actual filling factor may be larger than the nominal values derived with the least-squares fitting. In an extreme case, we estimate the magnetic flux by assuming $f=1$: the total \emph{vertical} magnetic flux then becomes $9.9 \times 10^{21}$ Mx, and the total \emph{horizontal} magnetic flux also becomes $2.0 \times 10^{22}$ Mx. The difference is a factor of 0.22 and 0.2, which roughly correspond to the average filling factor for the vertical and horizontal fields, respectively. If the contribution from stray-light reaches 50\% \citep{dos07}, the total \emph{vertical} flux is estimated to be $7.2 \times 10^{21}$ Mx. Note that we have to be careful in comparing the horizontal magnetic flux with the vertical flux due to a different sensitivity in the degree of Stokes Q, U and V polarization to transverse and line-of-sight magnetic field.

Since the inversion was performed for only 10.5\% of pixels with high S/N ratio, and the horizontal field strength is generally smaller than the vertical, we also obtained the horizontal magnetic flux from the wavelength-integrated Stokes V signals using a weak-field approximation. This would be less sensitive to error. We excluded pixels with flux $< 3 \sigma$ (7.2 Mx cm$^{-2}$), performed foreshortening correction in the pixel size, and added a correction (a factor of $\sqrt{2}$) for the transverse horizontal fields unseen in Stokes V. Total horizontal flux thus obtained is $4.35 \times 10^{21}$ Mx, which agrees well with the value obtained above.

The total \emph{vertical} magnetic flux for the whole area with latitude above $70\,^{\circ}$ is then estimated to be $5.6 \times 10^{21}$ Mx (with the filling factor $f$ considered) and $2.5 \times 10^{22}$ Mx (with $f=1$ assumed), assuming that the unobserved polar region has the same magnetic flux as that observed with SOT. An estimate considering the effect of stray-light is $1.8 \times 10^{22}$ Mx. Here, we chose only the flux tubes vertical to the local surface, and the horizontal flux is not included. Since the surface area with latitude above $70\,^{\circ}$ is $1.8 \times 10^{21}$ cm$^{2}$, the average flux is 3.1 G (with the nominal filling factor $f$ considered), 13.9 G (with $f=1$ assumed), and 10.0 G (with 50\% of stray-light taken into account). Though these are the most accurate flux estimation so far made for the polar regions, these number should be regarded as minimum values due to the threshold in the selection of pixels for accurate inversion.

\section{kG magnetic patches and acceleration of the fast solar wind}
\subsection{Comparison between photospheric and interplanetary magnetic flux}
The \emph{Hinode} X-ray image taken on March 16, 2007 shows that apparent polar coronal hole extends down to $60\,^{\circ}-\,70\,^{\circ}$ in latitude. Thus, the entire region shown in Figure 3 is the photospheric base of the polar coronal hole. We compare the magnetic flux in the polar photosphere and that observed in the interplanetary space in a different solar cycle. The mean magnetic field strength as observed with \emph{Ulysses} in 1993-1997 is 2.83 nT ($2.83 \times 10^{-5}$G) above $36\,^{\circ}$ heliolatitude at 1AU \citep{mc00}, and the total magnetic flux of the polar coronal hole is estimated to be $2\times10^{22}$ Mx, which is somewhat larger than the total photospheric magnetic flux obtained with the effective filling factor, and is close to the flux with the unit filling factor. These numbers are considered to be consistent, since (1) actual filling factor must be between one and the effective filling factor derived above, (2) there may be smaller undetected vertical flux tubes as indicated by the presence of spicules, (3) both measurements were done in different solar cycles, and (4) we obtained the photospheric flux with latitude higher than $70\,^{\circ}$, though the polar coronal hole extends latitude $<70\,^{\circ}$. 

\subsection{Polar flux tubes and the fast solar wind}

The fast solar winds emanate from the polar regions \citep{kr73, w97}. The vertical flux tubes would have a large expansion between the photosphere and the lower corona due to their high field strength, mono-polarity, and their very limited number and size in the polar region. The total area $S$ of the vertical flux tubes with average field strength $B \times f$ larger than 200 G ,which is identified with red contours in figure 4, is  $2.1 \times 10^{18}$ cm$^{2}$, and the total surface area of the photosphere corresponding to the SOT field of view is $7.2 \times 10^{20}$ cm$^{2}$. Thus, the areal expansion of individual flux tubes between the photosphere and the lower corona may reach a factor of 345.

The mean number density and velocity of the fast solar wind as observed by \emph{Ulysses} is 2.7 cm$^{-3}$ and 760 km s$^{-1}$, respectively at 1AU and at $>60\,^{\circ}$ heliolatitude. These values exhibit little variation with heliolatitude \citep{mc00}. On the basis of this, we estimate the total mass loss of $2.3 \times 10^{8}$ kg s$^{-1}$ from one of the polar regions, assuming uniform plasma parameters. The plasma density $\rho$ at $\tau_{5000} = 1$ is $3\times10^{-7}$ g cm$^{-3}$, and the upward speed associated with a fast solar wind is estimated to be only 2 cm s$^{-1}$. The apparent Doppler velocity further decreases due to projection effect. Indeed, we do not see any velocity feature at the locations of the vertical flux tubes.   

In a more mixed-polarity region, a larger fraction of the field lines will return at lower heights, allowing larger expansion for the ones that are indeed open higher in the corona. But, in the polar region, as soon as vertical field lines reach the chromosphere or the chromospheric-coronal boundary, the fields will expand since there is no obstacle for lateral expansion of the vertical flux tubes. Horizontal fields, though ubiquitous, would not reach the corona. All the open field lines forming the polar coronal hole essentially originate from such scattered small but intense magnetic patches, and the fast solar winds emanate from these vertical flux tubes seen in the photosphere (magnetic funnels, Tu \emph{et al.} 2005).

Alfv\'en waves are believed to play a vital role in the acceleration and heating of the fast solar winds \citep{jvh72, si06}. Alfv\'en speed rapidly increases with height due to the decrease in plasma density. A long-standing problem is that Alfv\'en waves with wavelength shorter than the Alfv\'enic scale height tend to be reflected back \citep{ai89, rlm91}. Rapid decrease in the magnetic field strength associated with rapidly expanding flux tubes near chromospheric boundary would make the vertical change in  Alfv\'en speed smaller, resulting in longer Alfv\'enic scale height. Therefore, the Alfv\'enic cutoff frequency may be lower in the polar flux tubes. We thus conjecture that the Alfv\'en waves generated in the photosphere are more efficiently propagated to the corona through the fanning-out flux tubes. Thus, the flux tubes may serve as the chimneys providing the entire coronal hole with Alfv\'en waves that accelerate solar winds.  

\section{Discussions}  
We discovered that the poloidal field near the pole has a form of unipolar flux tubes scattered in the polar region rather than a weak extended field. If the polar field with the same total magnetic flux $\Phi \sim BfS$ is uniformly distributed ($S$ is the total magnetic area), the estimated effective field strength would be about 10G as described above. The total magnetic energy is then proportional to $B^{2}fS = B\Phi$. Thus, the surface poloidal magnetic energy is approximately 90 times larger than the case for the uniform magnetic field, if we take $B \sim 900$G, corresponding to the peak of the energy PDF in Figure 5 (b). The equi-partition field strength $B_{e}$ is the field strength where magnetic energy is equal to kinetic energy of surface granular motion: $B_{e}=\sqrt{4 \pi \rho v^{2}}$. The typical equipartition field strength $B_{e}$ is about 400 Gauss for granules with a velocity of $v = 2 \times10^{5}$ cm s$^{-1}$ and $\rho$ given in section 4.2. The magnetic field strength for the majority of patches is larger than the equi-partition field strength. 
 
The observed unipolar strong flux tubes scattered in the polar region are considered to represent seed poloidal fields for toroidal fields \citep{wns89a, wns89b}. Magnetic flux is transported to the polar regions with meridional flows and supergranular diffusion in the flux-transport dynamo model \citep{dc99}. Since magnetic field has a form of such isolated flux tubes with super equi-partition strength instead of diffuse weak mean-field assumed in the flux transport dynamo \citep{dc99}, flux transport on the sun would be done via an aerodynamic (drag) force against magnetic tension force, and may be more difficult than the case for the mean field case assumed in the models. 

If the flux tubes seen on the surface of the Sun are maintained inside the Sun, this would affect a known difficulty in $\Omega$-mechanism \citep{wme56} to generate intense toroidal field: smaller amplification factor is needed to generate the same toroidal field from the poloidal field with intrinsic field strength of 1kG than from the averaged 10G field, and thus may be achievable within a solar cycle. We, however, recognize that there would remain a serious energetic problem, if the toroidal field strength indeed reaches 100kG \citep{sc96, mr06}.

Total flux of vertical magnetic field at the polar region estimated here is at most $7.2 \times 10^{21}$ Mx at the solar minimum, while various measurements on the total magnetic flux of single active region indicate $\sim 10^{22}$Mx \citep{lc07, jc07, mt08}. Thus, the measured total polar flux barely corresponds to that of single active region. The total toroidal flux would increase with time during the winding-up process by differential rotation, and the concept of the $\Omega$-mechanism would be viable with the observations presented here. 

The transient horizontal magnetic field discovered in the polar region appears to have properties similar to those found in quiet Sun and in active regions \citep{bwl07, rc07, os07, ri07, ri08}. In particular, PDFs of magnetic field strength for the polar region (Figure 5 (a)), quiet Sun, and active regions \citep{ri08} are remarkably similar, suggesting a common local dynamo process \citep{fc99} taking place all over the Sun. 

The X-ray telescope and EUV imaging spectrometer aboard \emph{Hinode} observed remarkable activity in the polar regions in a form of micro flares and jets \citep{sa07, ci07}. The lateral spreading of the vertical flux tubes to large area may be located well above the formation height of the two Fe lines, since there is no clear positional correlation between the horizontal fields and the vertical fields as seen in Figure 4. These X-ray jets could be due to magnetic reconnection at the lateral magnetic contacts with the horizontal fields and/or transient emergence of separate bipolar field lines \citep{ssh98, sh92}.

In conclusion, the magnetic landscape of the polar region is characterized by vertical kG patches with super equi-partition field strength, a coherency in polarity, lifetime with time scale of $5\,-\,15$ hours, and the ubiquitous weaker transient horizontal fields. The life time of the magnetic concentrations in the quiet Sun is estimated to be 2500 s for $2.5 \times 10^{18}$Mx, and 40 ks for $10 \times 10^{18}$Mx with \emph{MDI} \citep{hage99}. It is important to clarify similarities and differences between the polar region and the quiet Sun with \emph{Hinode}. This will be discussed in our subsequent paper.

\acknowledgments
S. T. thanks T. Magara, M. Rempel, K. Fujiki, T. Rimmele, T. J. Okamoto, N. Narukage, and R. Ishikawa for fruitful discussions. The authors express sincere thanks to the ISAS/JAXA \emph{SOLAR-B} launch team headed by Y. Morita for their exceptional achievement. Hinode is a Japanese mission developed and launched by ISAS/JAXA, collaborating with NAOJ as a domestic partner, NASA and STFC (UK) as international partners. This work was carried out at the NAOJ Hinode Science Center, which is supported by the Grant-in-Aid for Creative Scientific Research "The Basic Study of Space Weather Prediction" from MEXT, Japan (Head Investigator: K. Shibata), generous donations from Sun Microsystems, and NAOJ internal funding. The FPP project of LMSAL and HAO is supported by NASA contract NNM07AA01C.

\begin{figure}
\begin{center} 
\includegraphics[angle=0,width=7cm]{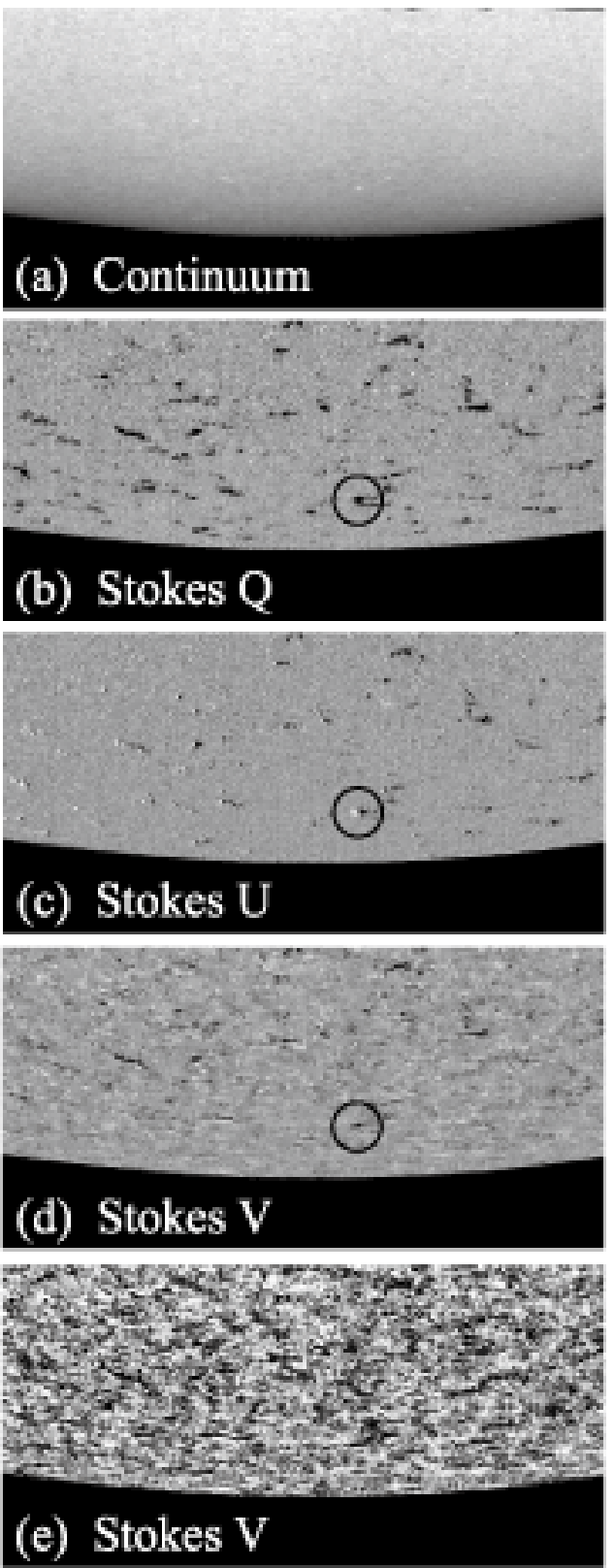}
\end{center}
\caption{
Polarization maps of the South polar region taken  at 12:02:19-14:55:48 on March 16 2007: 
\textbf{(a)} continuum map,
\textbf{(b)} Stokes Q map (transverse magnetic field),
\textbf{(c)} Stokes U map (transverse magnetic field),
\textbf{(d), (e)} Stokes V map (line-of-sight magnetic field).
East is to the left, and North is up. The gray scale indicates the wavelength-integrated total degree of polarization. The displayed images saturate at the degree of polarization  $+/-$ 0.003 for panels \textbf{(b)}, \textbf{(c)}, and \textbf{(e)}, while at  $+/-$ 0.02 for panel \textbf{(d)}. Black circles in panels \textbf{(b)}, \textbf{(c)}, and \textbf{(d)} indicate a fanning-out magnetic structure.}   
\end{figure}
\clearpage

\clearpage
\begin{figure}
\begin{center} 
\includegraphics[angle=90,width=10cm]{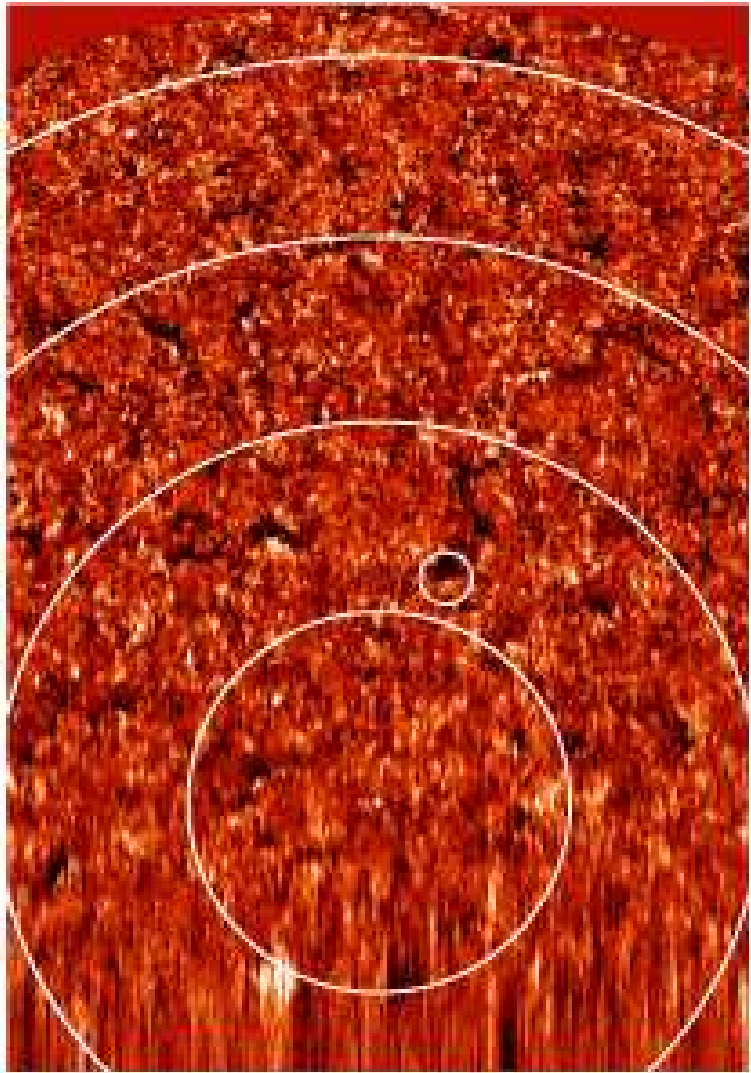}
\end{center}
\caption{
Polar view of the circular polarization (Stokes V) signal created from figure 1(d) and (e). The original observing field of view (Figure 1) is 327".52 (East-West) $\times$ 163".84 (North-South), and it is converted to the map seen from above the South Pole. For orientation, left on the page is east, right is west, and the observation was carried out from top of the page. The spatial resolution is lost near the extreme limb (i.e. near the bottom of the figure). The field of view is 327".52 (East-West) $\times$ 472".96 (North-South along line of sight). The field of view for the line-of-sight direction (163".84) is expanded to 472".96 due to foreshortening correction. Pixel size is 0".16. Latitudinal lines for $85\,^{\circ}$, $80\,^{\circ}$, $75\,^{\circ}$, $70\,^{\circ}$ are shown as white circles, while the cross mark indicates the South Pole. 
The displayed images saturates at the degree of polarization $+/-$ 0.005. Black indicates the magnetic field component toward the observer, while white away from the observer. A circle indicates the same magnetic patch indicated by circles in Figure 1.}
\end{figure}
\clearpage

\begin{figure}
\begin{center}
\includegraphics[angle=90,width=10cm]{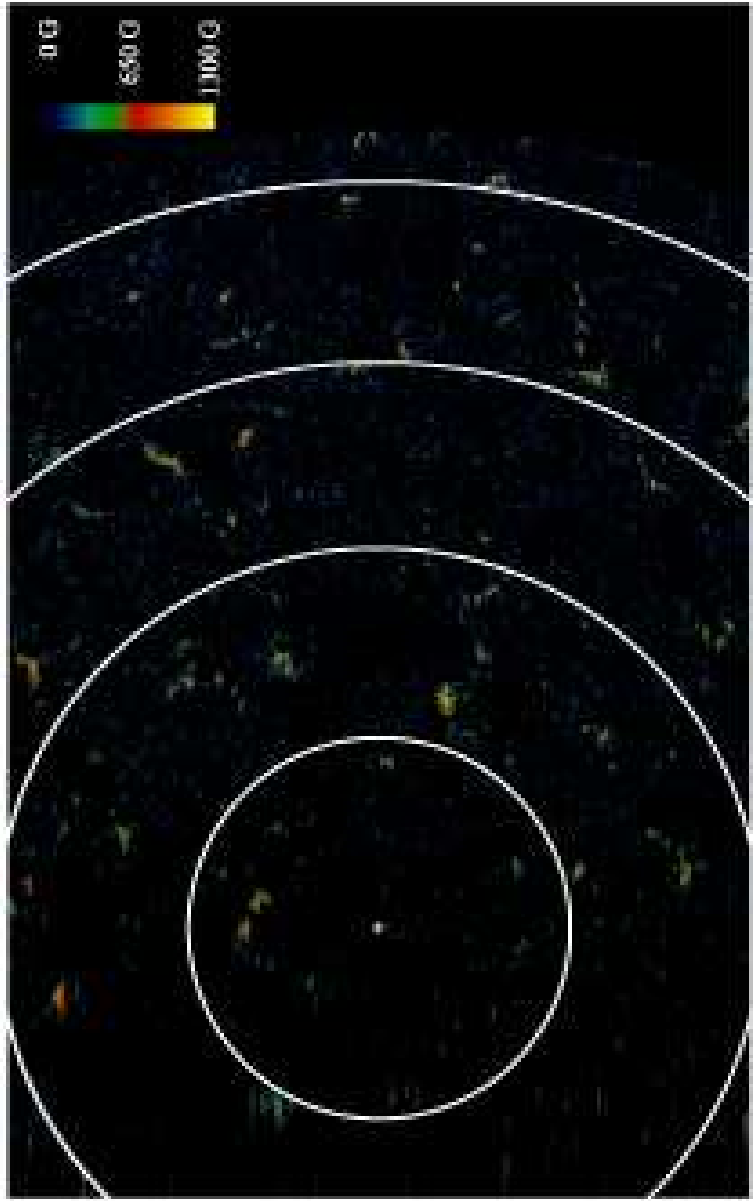}
\end{center}
\caption{South polar view of the magnetic field strength taken at 12:02:19$\,-\,$14:55:48 on March 16 2007. The captions for the polar view are the same as Figure 3. Magnetic field strength is obtained for pixels meeting a given threshold (see text).}
\end{figure}
\clearpage

\begin{figure}
\begin{center}
\includegraphics[angle=90,width=10cm]{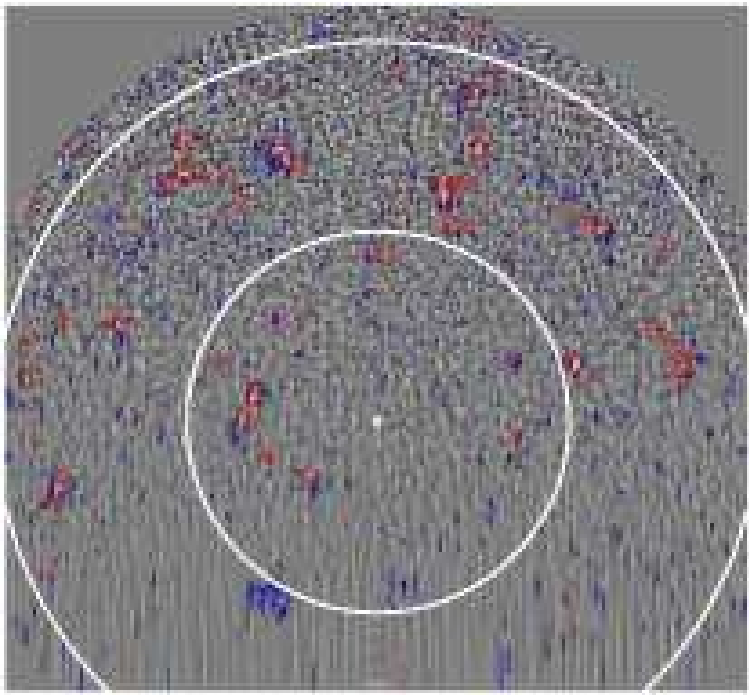}
\end{center}
\caption{Polar view in the continuum for latitude higher than $80\,^{\circ}$ created from figure 1(a). Contours indicate locations with average field strength $B \times f$ of 200G. (A region inside the contour has average field strength larger than 200G.) Red contours indicate regions where the local inclination $i$ smaller than than $25\,^{\circ}$ (vertical), while blue contours show regions with local inclination larger than $65\,^{\circ}$ (horizontal). East is to the left, and West is to the right. Latitudinal lines for $85\,^{\circ}$ and $80\,^{\circ}$ are shown, with a cross mark indicating the South Pole. Near the extreme limb (left side), spatial resolution is lost.}
\end{figure}

\newpage      
\begin{figure*} [htps]
\begin{center}
\includegraphics[width=15cm, angle=0]{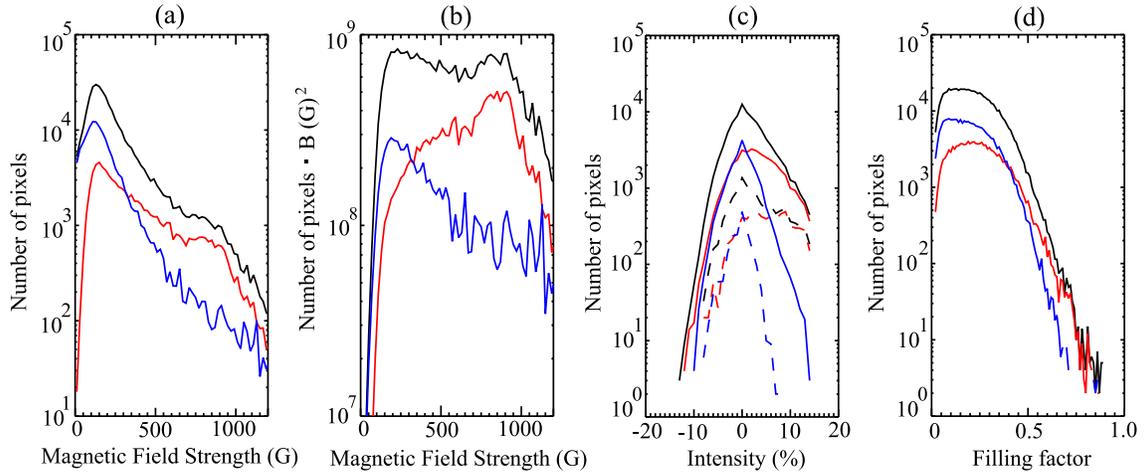}
\end{center}
\caption{
(a) Number of pixels as a function of the magnetic field strength (probability distribution function, PDF, black line). Red lines indicate vertical field, blue horizontal field, and black total for all the panels. (b) Number of pixels $\times B^2$ as a function of the magnetic field strength. The bin size of magnetic field strength in panels (a) and (b) is 20G. (c) Histogram of continuum intensity with magnetic field strength $>300$G (solid line) and $>800$G (short broken line). Since continuum intensity rapidly decreases toward the limb, the horizontal axis is the normalized excess continuum level with respect to the continuum level averaged over a 6".4 box.  (d) Filling factor as defined in section 3.1. All the panels are for latitudes $>75\,^{\circ}$. 
}
\end{figure*}

\end{document}